%% file: ecir2026-querysim.tex
\documentclass[runningheads]{llncs}

\usepackage[T1]{fontenc}
\usepackage{url}

\usepackage{tabularx}
\usepackage{multirow}
\usepackage{color}
\usepackage{booktabs}
\usepackage[square,comma,numbers,sort&compress,sectionbib]{natbib}
\usepackage{enumerate}
\usepackage{graphicx}
\usepackage{enumitem}
\usepackage{makecell}
\usepackage{comment} %

\usepackage[table,xcdraw]{xcolor} %
\usepackage{tabularx}
\usepackage{booktabs}

\begin{document}

\title{Validating Search Query Simulations: A~Taxonomy of Measures}

\author{Andreas Konstantin Kruff\inst{1}\thanks{Equal contribution.}\orcidID{0009-0002-8350-154X} \and
Nolwenn Bernard\inst{1}\textsuperscript{\(\star\)}\orcidID{0009-0007-0565-3210} \and
Philipp Schaer\inst{1}\orcidID{0000-0002-8817-4632}}

\authorrunning{Kruff et al.}

\institute{TH Köln - University of Applied Sciences, Germany\\ 
\email{firstname.lastname@th-koeln.de}}
\maketitle              
\begin{abstract}

Assessing the validity of user simulators when used for the evaluation of information retrieval systems remains an open question, constraining their effective use and the reliability of simulation-based results. To address this issue, we conduct a comprehensive literature review with a particular focus on methods for the validation of simulated user queries with regard to real queries. Based on the review, we develop a taxonomy that structures the current landscape of available measures. We empirically corroborate the taxonomy by analyzing the relationships between the different measures applied to four different datasets representing diverse search scenarios. Finally, we provide concrete recommendations on which measures or combinations of measures should be considered when validating user simulation in different contexts.
Furthermore, we release a dedicated library with the most commonly used measures to facilitate future research.

\end{abstract}

\input{ecir2026-querysim-01}
\input{ecir2026-querysim-02}
\input{ecir2026-querysim-03}
\input{ecir2026-querysim-04}
\input{ecir2026-querysim-05}
\input{ecir2026-querysim-06}

\begin{credits}
\subsubsection{\ackname}
This work is partially funded by Deutsche Forschungsgemeinschaft (DFG) under grant number 509543643 and within the funding programme FH-Personal (PLan CV, reference number 03FHP109) by
the German Federal Ministry of Education and Research (BMBF) and Joint Science Conference (GWK).

\subsubsection{\discintname}
The authors have no competing interests to declare that are relevant to the content of this article.
\end{credits}

\input{ecir2026-querysim-appendix}

\bibliographystyle{splncs04nat}
\bibliography{ecir2026-querysim.bib}

\end{document}

%% file: ecir2026-querysim-01.tex
\section{Introduction}
\label{sec:intro}

Validation of user simulators remains one of the main blocking points for their wide adoption in the field of (interactive) information retrieval~\citep{Breuer:2025:SIGIRForum}. Indeed, validation contributes to building trust in the simulation and its results. To the best of our knowledge, there are no standardized methodology and measures to validate user simulators. Validation is a faceted problem, as previous work show different interpretations of validation~\citep{Balog:2024:FnTIR,Zeigler:2019:book}, which may explain this absence of a unified framework. For example, \citet{Breuer:2022:ECIR} validate simulation with regards to retrieval performance, characteristics of simulated search sessions, and term-based similarity, while \citet{Huurnink:2010:CLEF} validate simulation by comparing system rankings based on performance metrics. 

Simulation in information retrieval can focus on different parts of the search process, for example, query variants generation~\citep{Alaofi:2023:SIGIR,Alaofi:2025:ICTIR,Breuer:2022:ECIR} and clicks~\citep{Chuklin:2015:book}. In this work, we specifically focus on query simulation and its validation. Query simulation aims to produce queries mimicking real user queries that can be used to generate synthetic data for training and evaluating information retrieval systems in a scalable and reproducible manner. Despite the importance of validation and the long existence of the field, to the best of our knowledge, a comprehensive overview of validation facets and measures for the query simulation task is still missing. Hence, the choice of validation facets and measures often remains ad-hoc and based on intuition given the specific context of the simulation.

In this work, we provide a comprehensive overview of validation facets and associated measures in the form of a taxonomy based on a literature review.
This taxonomy provides a starting point for practitioners to select the facets and measures that are relevant for their simulation approach and the available data.
The taxonomy is built using a bottom-up approach, starting from the measures used in previous work, we group them into common facets.
The taxonomy is at most four levels deep, providing fine-grained facets divided between two main meta-facets: (1) facets related to the inability to distinguish between simulated and real data, and (2) facets related to performance prediction capabilities of a simulator.
To corroborate the identified facets, we investigate the relationships between the different measures in multiple use cases. For this analysis, we look at correlations and mutual information between the measures. It aims to identify potential complementarities and redundancies between the measures, intuitively guiding the selection of measures for future work. Across the datasets, the correlations were predominantly linear and monotonic. Only one dataset showed notable non-linear relationships, based on the applied thresholds. The exploratory factor analysis generally indicated two to three underlying factors, with the measures loading consistently on these factors.

In summary, the contributions of this work are threefold. First, we propose a taxonomy of validation facets and associated measures for query simulation. Second, we provide an analysis of the relationships between the measures, which inherently reflects the relationships between facets. Third, we release a library to automatically compute measures from the taxonomy, facilitating future research in the field.

%% file: ecir2026-querysim-02.tex
\section{Literature Review}
\label{sec:review}

This work aims to provide a general overview of previous work on validation approaches applied to query simulation in the context of (interactive) information retrieval.
Therefore, we ask the following research questions: 
\begin{description}
    \item[\textbf{RQ1:}] What are the different facets considered when validating query simulation in (interactive) information retrieval?
    \item[\textbf{RQ2:}] What are the measures associated with these facets?
\end{description}

To answer these questions, we review the literature, as described in Section~\ref{sec:review:method}, on query simulation with a focus on validation approaches. The findings of this review (Section~\ref{sec:review:results}) serve as a foundation for our proposed taxonomy of validation facets and measures, which is presented in Section~\ref{sec:taxonomy}.

\subsection{Methodology}
\label{sec:review:method}

For this literature review, we select four source databases: ACM Digital Library,\footnote{\url{https://dl.acm.org/}} IEEE Xplore,\footnote{\url{https://ieeexplore.ieee.org/}} Springer Nature Link,\footnote{\url{https://link.springer.com/}} and arXiv.\footnote{\url{https://arxiv.org/}} Our choice is mainly motivated by the research areas indexed by these sources, i.e., the first two sources index papers from computer science, while the last two sources also index papers in other areas~\citep{Gusenbauer:2020:ResSynMeth}.
To have a broad coverage of the literature, we design a search query that captures papers related to query simulation in general. Therefore, we use the following query over all the different fields of the sources: \texttt{"query simulation" OR "simulated quer*"}.
Note that the wildcard `*' operator is not supported by arXiv, so we adapt the query to: \texttt{"query simulation" OR "simulated queries" OR "simulated query"}. 

\subsection{Results}
\label{sec:review:results}

\begin{table}[t]
    \centering
    \caption{Overview of the results from the literature review with digital libraries.}
    \label{tab:lit_review}
    \begin{tabular}{l c c c}
        \toprule
        \textbf{Source} & \textbf{\# Papers} & \textbf{\# Relevant} & \textbf{Papers incl. validation} \\ \midrule
        ACM Digital Library & 66 & 12 & \citep{Labhishetty:2022:ICTIR,Azzopardi:2007:SIGIR,Alaofi:2023:SIGIR,Alaofi:2025:ICTIR,Carterette:2015:ICTIR,He:2025:SIGIR,Elsweiler:2025:SIGIR,Labhishetty:2021:SIGIR,Elsweiler:2011:SIGIR}  \\
        IEEE Xplore & 26 & 5 &  \citep{Traub:2016:JCDL,Morrison:2011:IEEE,Cai:2009:IEEE}\\
        SpringerLink & 82 & 8 & \citep{Breuer:2022:ECIR,Engelmann:2024:ECIR,Zerhoudi:2022:TPDL,Sinha:2024:ECIR,Berendsen:2012:CLEF,Huurnink:2010:CLEF,Ben:2005:ECIR} \\
        arXiv & 26 & 2 & \citep{Erbacher:2022:SIGIR} \\ \midrule
        \textbf{Total} & 200 & 27 & 20 \\ \bottomrule
    \end{tabular}
\end{table}

The search query was executed on the four sources on August 1, 2025, and returned a total of 200 papers (including duplicates). To keep the papers relevant to our research questions, we applied a two-step filtering process performed by two of the authors independently; uncertainties were resolved through discussion.\footnote{The list of annotated papers is made public at: \url{http://bit.ly/4sxDYWT}.}
First, we filtered the papers based on their title, abstract, and introduction to ensure they were related to query simulation in the context of (interactive) information retrieval. Second, after reading the full text of the 27 remaining papers, we filtered out those that did not include any form of validation of the simulation. At the end of this process, 20 papers are kept for further analysis, as summarized in Table~\ref{tab:lit_review}. Additionally, we also added 4 papers~\citep{Gunther:2021:Sim4IR,Zhang:2025:SIGIR,Zendel:2025:CIKM,Rahmani:2025:CIKM} that were not retrieved by the search query but were known to us and matched the selection criteria. The final set of selected papers is analyzed in the next section to identify the different facets and measures used to validate query simulation.

%% file: ecir2026-querysim-03.tex
\section{Taxonomy of Validation Facets and Measures}
\label{sec:taxonomy}

\begin{figure}[t]
    \centering
    \includegraphics[width=\textwidth,keepaspectratio]{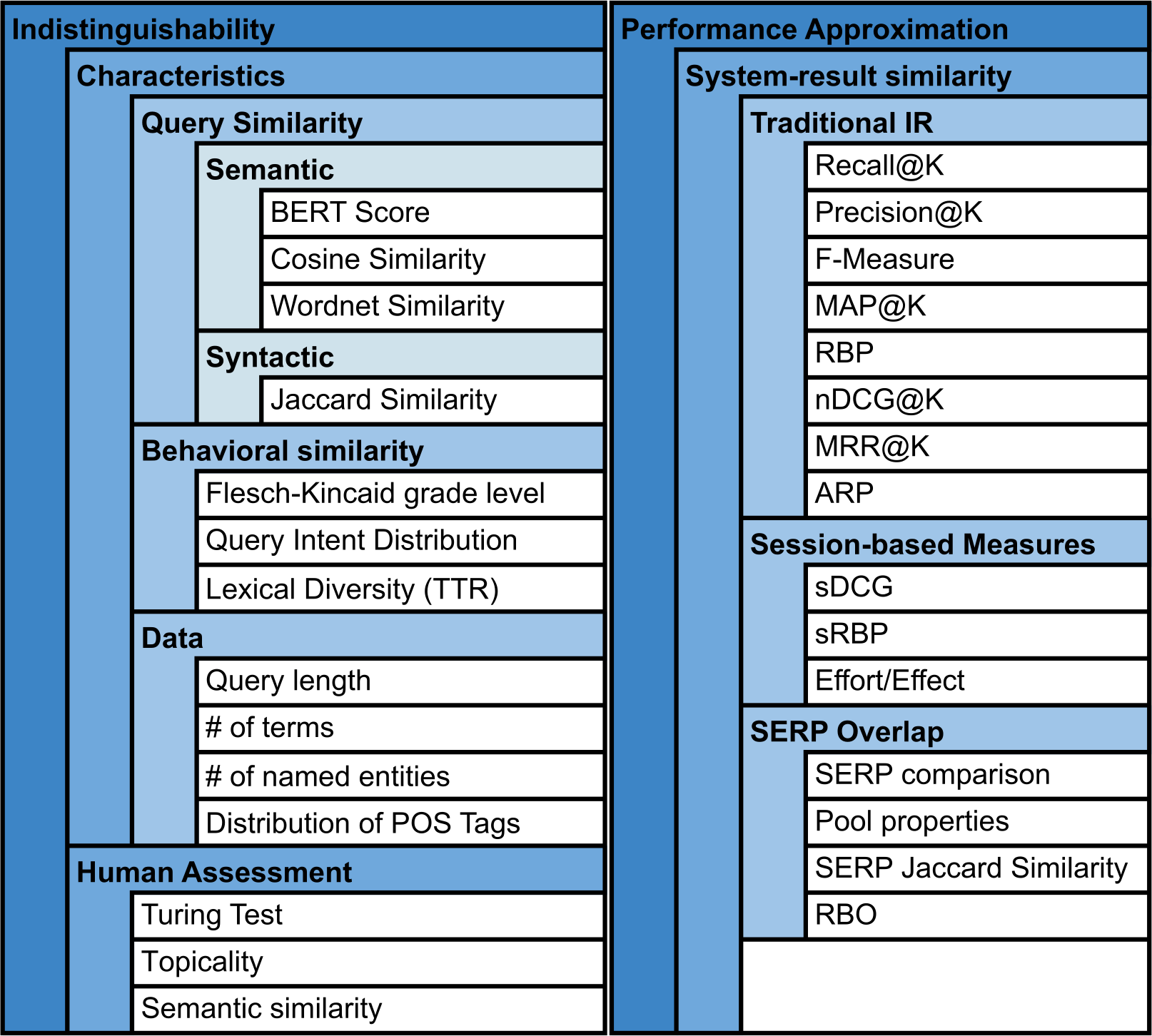}
    \caption{Taxonomy of validation facets (in blue) and measures (in white) for query user simulation.}
    \label{fig:taxonomy}
\end{figure}

The 24 papers selected are analyzed with regard to the validation facets and associated measures they consider. Following a bottom-up approach, we iteratively group the specific facets into increasingly abstract categories, resulting in higher-level nodes that capture broader validation dimensions. This process leads to the taxonomy of validation facets and measures presented in Figure~\ref{fig:taxonomy}. 
We observe two main meta-facet in the taxonomy, one focusing on the \emph{indistinguishability} of the simulated data from real user data and the other focusing on the \emph{performance approximation} abilities of the simulation.
We argue that the former tends to promote a user-centric perspective for the validation, while the latter is more system-centric.
Indeed, we find that \emph{indistinguishability} can be assessed automatically with regards to specific characteristics such as the similarity between the simulated and real queries, e.g.,~\citep{Alaofi:2023:SIGIR,Labhishetty:2022:ICTIR,Breuer:2022:ECIR,Zhang:2025:SIGIR} or the structure of the queries, e.g.,~\citep{Traub:2016:JCDL}. Furthermore, it can also be assessed by human evaluation, for example, by asking human judges to distinguish between search sessions with real and simulated queries~\citep{Gunther:2021:Sim4IR} or to evaluate the semantic similarity of the simulated queries to human queries~\citep{Alaofi:2025:ICTIR}.
On the other hand, \emph{performance approximation} is more concerned with the analysis of the outcome produced by the simulated queries, i.e., are the results obtained similar to the ones obtained with real queries? A common approach to answer this question consists of computing performance metrics, such as traditional information retrieval metrics, for the simulated queries and the real queries with one or more systems; then, comparing the proximity of the results or, in the case of multiple systems, the ranking of the systems based on the simulated queries to the ranking based on real queries corresponding to a tester-based approach~\citep{Labhishetty:2022:ECIR}. For example, \citet{Zerhoudi:2022:TPDL} use mean square logarithmic error to assess the proximity between isoquants of simulated and real queries, while, \citep{Huurnink:2010:CLEF,He:2025:SIGIR} look at the correlation between mean reciprocal rank-based rankings of the systems based on real and simulated queries.
Another approach is to analyze the overlap between the search engine results pages (SERPs) produced for the simulated and real queries. For example, \citet{Traub:2016:JCDL} analyse the overlap between the retrieved documents for the simulated and real queries.
We acknowledge that additional measures have been used in the reviewed literature, but are not represented in the taxonomy. This choice is motivated by the fact that these measures are either highly domain-specific or have not been used in multiple papers, indicating their limited adoption to multiple application scenarios. It includes the average NewsGuard score~\citep{Elsweiler:2025:SIGIR} and the Earth Mover Distance between Tip-of-the-Tongue linguistic codes~\citep{He:2025:SIGIR}. 

We also observe that statistical comparison is often used in complement to the other measures. For example, correlation coefficients are often used when comparing rankings of systems based on real and simulated queries, e.g.,~\citep{Breuer:2022:ECIR,He:2025:SIGIR,Sinha:2024:ECIR,Berendsen:2012:CLEF}. Additionally, statistical significance tests can be employed to assess if simulated and real data follow the same distribution (indicating indistinguishability). 
Statistical comparison can strengthen the validation by showing that the results are not due to chance.

We note that not all the facets and measures proposed in the taxonomy are necessarily applicable to all query simulation approaches. This is particularly the case for the \emph{performance approximation} facet, for which some of the measures require relevance judgments, e.g., normalized discounted cumulative gain or mean average precision. However, we argue that the taxonomy can serve as a starting point to decide which facets and measures can be used to validate a specific query simulation approach. Furthermore, it can easily be extended with additional facets and measures, including some from other fields, such as natural language generation.

%% file: ecir2026-querysim-04.tex
\section{Experimental setup}
\label{sec:case_study}

In order to support the proposed taxonomy of validation facets and measures, we conduct a comprehensive analysis of the relationships between different measures. It allows us to investigate to what extent the various measures complement each other in their informational value or overlap due to redundant information. The intuition is that measures belonging to the same facet should share similar information, while measures from different facets should capture complementary aspects of the simulation. 

This section first describes the method applied for this analysis (Section~\ref{sec:exp:method}) and then presents the datasets used for the experiments (Section~\ref{sec:exp:datasets}). 

\subsection{Methodology}
\label{sec:exp:method}

We propose a methodology in five steps to analyze the relationships between different measures in our taxonomy and test our initial intuition regarding the facets. The first step is optional, depending on the data and resources available. We describe each step as follows:
\begin{enumerate}[start=0]
    \item (optional) \textbf{Augment data} with search engine results pages (SERPs) given that a retrieval system with the document collection from the original dataset indexed is available. This step allows the computation of performance prediction measures, especially those related to the analysis of overlap between the results obtained with the simulated and real queries (e.g., SERP Jaccard, RBO).
    \item \textbf{Compute measures} from the taxonomy that are applicable given the available data. Measures could be computed for each pair of simulated and real queries (one-to-one) or for a set of simulated queries corresponding to a real query (one-to-many). In this work, we consider only one-to-one comparisons to ensure consistency across the datasets. 
    \item \textbf{Conduct exploratory factor analysis (EFA)} on the computed measures to identify underlying latent factors. We inspect the factor loadings to see if measures that load highly on the same factor correspond to the same facet in the taxonomy.
    \item \textbf{Analyze correlation matrices} of the computed measures to identify linear and monotonic relationships. We compute Pearson $\rho$ and Kendall's $\tau$ correlation coefficients between all pairs of measures. The former captures linear relationships, while the latter captures monotonic relationships. We look at the strength and direction of the correlations to identify clusters of measures that share similar information and compare them to the facets in the taxonomy.
    \item \textbf{Analyze mutual information} between all pairs of measures to identify potential nonlinear dependencies. We compute normalized mutual information (NMI) to quantify the amount of shared information between two measures, regardless of the nature of their relationship. We look for pairs of measures with high NMI but low Pearson and Kendall correlations, as they may capture complementary aspects of the simulation not reflected in linear or monotonic relationships.
\end{enumerate}
Based on the results from steps 2--4, we consider two criteria to support our proposed taxonomy. First, the clusters of measures identified by EFA should mainly align with the facets in the taxonomy. Second, measures within the same facet should demonstrate high correlation and mutual information, while measures from different facets should show lower correlation and mutual information, indicating that they capture complementary aspects of the simulation. 

\subsection{Datasets}
\label{sec:exp:datasets}

\begin{table}[t]
    \centering
    \caption{Datasets overview.}
    \label{tab:datasets}
    \begin{tabular}{p{2.8cm}@{\hskip .8em} c@{\hskip .8em} c@{\hskip .8em} p{4cm}@{\hskip .8em} c}
        \toprule
        \textbf{Dataset} & \textbf{\makecell[c]{Queries}} & \textbf{\makecell[c]{Simulations}} & \textbf{\makecell[l]{Document collection}} & \textbf{\makecell[c]{Qrels}} \\ \midrule
        Sim4IA 2025 \citep{Kruff:2026:ECIR}  & 35 & 31 + 19 & CORE\textsuperscript{$\dagger$} & No \\
        UQV100~\cite{Alaofi:2023:SIGIR,Bailey:2016:SIGIR} & 100 & 3 & ClueWeb12-B\textsuperscript{$\ddagger$} &  Yes\\
        UQV subset~\citep{Breuer:2022:ECIR} & 50 & 21 & Common Core 2017~\citep{Allan:2017:TREC} & Yes \\
        DL seed queries~\cite{Alaofi:2025:ICTIR} & 126 & 19 & MS MARCO Passage v2~\citep{Bajaj:2018:arXiv} & Yes \\\bottomrule
        \multicolumn{5}{l}{\textsuperscript{$\dagger$} \footnotesize \url{https://core.ac.uk/}} \\
        \multicolumn{5}{l}{\textsuperscript{$\ddagger$} \footnotesize \url{https://www.lemurproject.org/clueweb12.php/}} \\
        \end{tabular}
\end{table}

We applied our proposed methodology on four public datasets (see Table~\ref{tab:datasets}):

\begin{itemize}
    \item \textbf{Sim4IA 2025}: 
    The Sim4IA dataset originates from the Sim4IA 2025 Micro-Shared Task Workshop~\cite{Schaer:2025:SIGIR}. It is based on original search sessions derived from CORE log files and provides one query per session as the gold standard for evaluation. Participants are instructed to generate a ranked list of 10 candidate queries for prediction (i.e., Task A). Consequently, the analysis can be carried out either one-to-one based on the ranking or one-to-many, averaged across all candidate queries per session.
    
    \item \textbf{UQV100}:
    The original UQV100 collection comprises 100 topics, each accompanied by a backstory and associated queries. Here, GPT-3.5 is used to generate query variants with temperatures of 0.0, 0.5, and 1. Since the number of variants produced per topic varied, the first generated variant is consistently selected to avoid skewing the evaluation measures~\citep{Alaofi:2023:SIGIR}.
    
    \item \textbf{UQV subset}:
    This dataset contains simulated query variants using the UQV collection from TREC Common Core 2017. It involves 21 simulators, each of which generated ten query variants, that are grouped into two methods, namely the TREC Topic Searcher (TTS) and the Known Item Searcher (KIS). In addition, eight modification strategies are applied that vary in the degree of term variation and selection~\citep{Breuer:2022:ECIR}. Furthermore, eight human annotators produced up to ten variants per original query. This setup enables multiple one-to-one and one-to-many comparisons between the human annotators and the simulators. Neither the human user queries nor the simulated queries are explicitly ranked; however, we implicitly assume that the first generated variant corresponds to the most natural reformulation.
    
    \item \textbf{DL seed queries}:
    This dataset is based on seed queries from the Deep Learning Tracks 2021 with 53 queries and 2022 with 73 queries. Query variants were generated with GPT-4 using a temperature of 1. The variants are created with different personas, user groups, and textual transformation strategies, as well as in a neutral setting without any persona. For each configuration, three variants are generated, but they are also not explicitly ranked~\citep{Alaofi:2025:ICTIR}.
\end{itemize}

The Sim4IA dataset is unique as there are no comparable datasets available that focus on next query prediction based on real user sessions using simulation. For further evaluation, we therefore rely on datasets derived from the UQV collections to apply and assess various query variant strategies. Although the underlying tasks differ, as contextual information for the predictions is missing, we argue that this should affect only the absolute quality of the results, but not the correlations between the measures. Due to the given topic descriptions or backstories, the simulators have different yet related contexts for the query simulation.
While the underlying test collection for the Sim4IA dataset unfortunately does not include relevance judgments, which prevents the calculation of traditional IR measures, the three other collections do provide such judgments. 

%% file: ecir2026-querysim-05.tex
\section{Analysis}
\label{sec:results}

In this section, we summarize the findings from the exploratory factor analysis (EFA), the correlation matrices analysis, and mutual information (i.e., steps 2--4 of the method). These different analyses reveal consistent patterns across the different datasets.

The EFA consistently identifies three to four main dimensions underlying query evaluation. Classical information retrieval (IR) performance metrics (i.e., nDCG, Precision, Recall, MAP, and MRR) are loaded together, confirming that they capture highly overlapping aspects of retrieval effectiveness and are largely redundant.
Query similarity measures (i.e., BERT Score, Jaccard, Cosine, and WordNet similarities) form a separate cluster, indicating a complementary perspective on conceptual overlap between the simulated and the real queries.
SERP overlap measures (i.e., SERP Jaccard similarity and RBO) tend to either form a distinct factor or cluster together with query similarity measures. This reflects their role in capturing similarity in result rankings, which is related to but not fully determined by query-level similarities.
Behavioral and descriptive query features (i.e, query length, number of unique terms, Fleisch-Kincaid grade level, lexical diversity, number of named entities) generally constitute a distinct, loosely defined factor.

\begin{figure}[t]
    \centering
    \includegraphics[width=\textwidth,keepaspectratio]{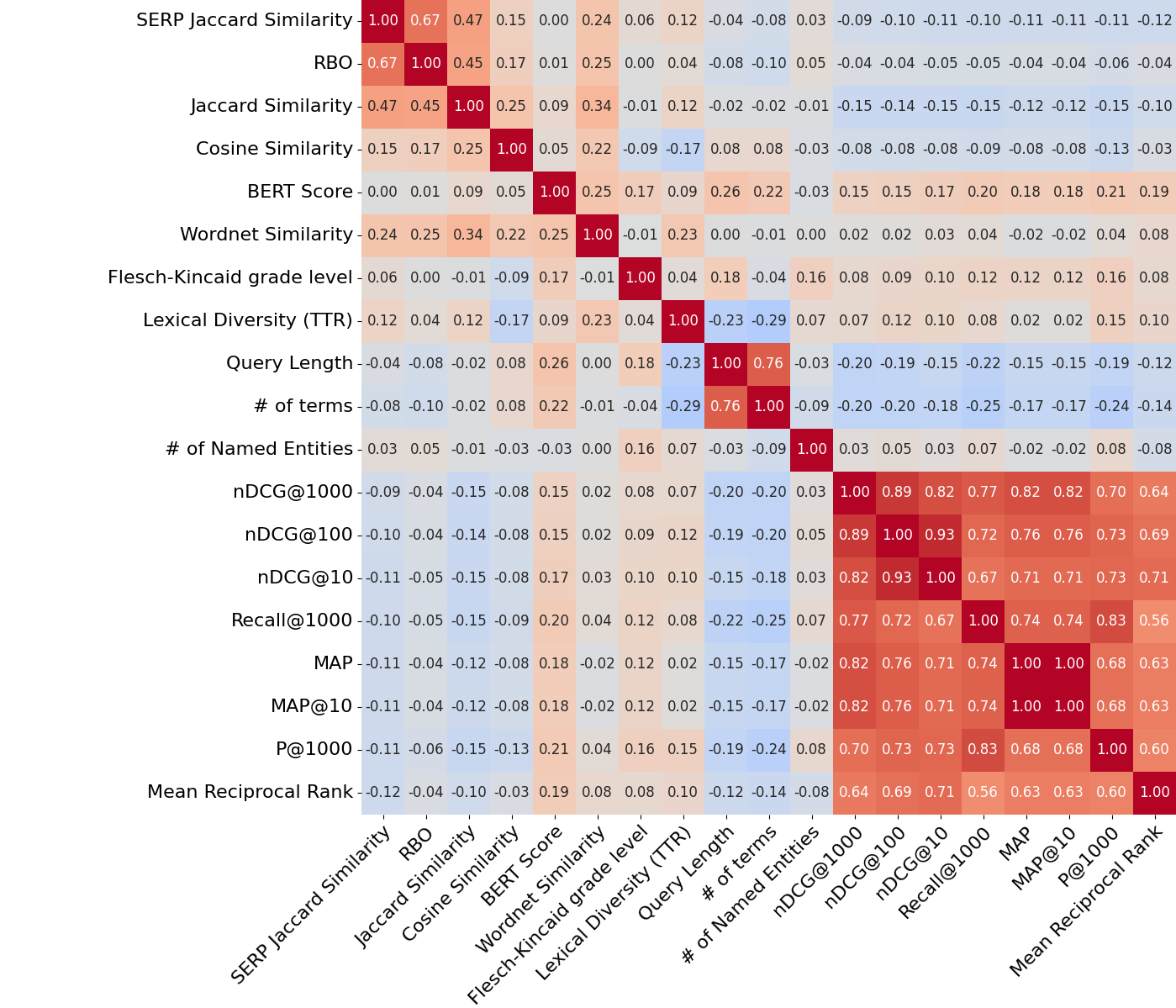}
    \caption{Pearson correlation matrix for the DL 2021 seed queries.}
    \label{fig:heatmap}
\end{figure}

The analysis of the correlation matrices across the datasets aligns with these findings.
Pearson correlation matrices indicate strong linear relationships among classical IR metrics ($\bar{\rho} = 0.77$) and basic query characteristics ($\bar{\rho} = 0.91$). The two SERP Overlap measures are also highly correlated across datasets ($\bar{\rho} = 0.85$). Query similarity measures exhibit a moderate internal correlation ($\bar{\rho} = 0.47$) and a similarly moderate correlation with SERP Overlap measures ($\bar{\rho} = 0.41$). The average correlation coefficients were obtained by first averaging over all measure pairs and then across the four datasets. Figure~\ref{fig:heatmap} shows a representative heatmap of the observed patterns, with the BERT score as a notable exception. Kendall’s $\tau$ correlation matrices largely mirror these patterns, though associations are slightly weaker. Mutual information analysis indicates no relevant nonlinear dependencies for most datasets. In the Sim4IA dataset, limited nonlinear associations are observed, mainly involving semantic similarity, lexical diversity, and Fleisch-Kincaid grade level.

Overall, the analysis indicates that while classical IR measures are highly redundant, query similarity measures and SERP overlap measures provide additional, complementary dimensions for evaluating simulated query validity. Nonlinear relationships appear to be rare and dataset-specific, reinforcing the dominance of linear and monotonic associations.

Unsurprisingly, the analysis of the measures across the four datasets reveals a consistently high correlation among traditional information retrieval (IR) measures. These measures are also the most frequently used in the studies included in our literature review, with many studies employing multiple traditional IR measures within the same study. This indicates that researchers place considerable value on system-based query performance similarity. However, our analysis suggests that using multiple traditional IR measures in this way does not provide substantial additional insights for evaluation purposes.

In contrast, SERP overlap measures, which are also categorized as ``performance approximation'' measures and reflect system-side evaluation, offer complementary information. Their predictive power is not redundant with that of traditional IR measures, and they tend to correlate more with query similarity measures. Although both clusters evaluate queries from a system perspective, we suggest using them together, as they capture different aspects of system performance. Furthermore, the partial correlation observed between SERP overlap and query similarity measures provides additional information about string-based query similarity.

Similarly, query similarity and SERP overlap measures show moderate correlation depending on the dataset. Nevertheless, these measures complement each other by capturing distinct dimensions of system ranking and query similarity. Within query similarity measures themselves, moderate intercorrelation exists, yet each measure carries unique information. Depending on the evaluation task, one might prioritize lexical similarity, semantic similarity, or a combination of both to achieve a more comprehensive assessment of query similarity.

For behavioral and descriptive data statistic measures, no consistent correlations could be observed across the datasets. Even within the clusters, correlations are generally negligible, except for query length and query number of terms, which show high correlation, as might be expected. This indicates that these measures capture distinct aspects of the query and user behavior that are not reflected in other clusters.

We acknowledge that restricting the analysis to a one-to-one comparison based on the top-1 simulated query per original query could potentially bias the observed correlations. To assess the robustness of our findings, we applied a bootstrapping procedure over 1,000 iterations, randomly selecting simulated queries referring to the same topic within the same simulator and across different simulators for each iteration. The results show that both Pearson and Kendall correlations exhibit only minor deviations, with maximum absolute differences of approximately 0.15 and the majority of deviations being substantially lower. These findings indicate that the choice of the specific simulated query within a topic has a negligible impact on the overall correlation analysis. Notably, measures based on query statistics often show relatively higher deviations in their correlations with other measures. This can be explained by the fact that semantic similarity measures (e.g., cosine similarity) are not necessarily related to query length, so depending on which simulated query is selected, correlations between length-dependent and semantic measures can fluctuate, even if the queries are semantically similar.

%% file: ecir2026-querysim-06.tex
\section{Conclusion}
\label{sec:conclusion}

In this paper, we investigate the methods and measures used to validate user query simulation approaches. To provide an overview of the current landscape of validation facets and measures, we propose a taxonomy based on a literature review and corroborate it with an empirical analysis of the relationships between the measures. All measures computed in this analysis are bundled into a complementary software library, allowing for reproducibility and further experimentation on the matter of query simulation validation.

While the taxonomy offers a practical conceptual framework, our analysis shows that the actual relationships are more nuanced and context-dependent. In particular, traditional IR performance metrics are found to be highly redundant, capturing largely overlapping aspects of retrieval effectiveness. In contrast, semantic similarity measures provide complementary information about conceptual overlap, and SERP-based overlap measures often capture unique patterns not reflected by either traditional or semantic metrics. 

We acknowledge that some measures included in the current taxonomy could not be evaluated in this study. Indeed, each measure has specific data requirements, and not all datasets provide the necessary data. For example, some measures require relevance judgments that are not available in all datasets, while others need human assessments that are costly and difficult to reproduce. 

Although we were unable to compare all available measures included in the taxonomy, we argue that the validation of user query simulation should consider different measures, preferably from both user- and system-centric facets, such as query similarity and traditional IR measures, to obtain a comprehensive assessment. A more fine-grained analysis or complementary user studies focusing on human assessment methods, such as Turing tests, are left as future work. 
Moreover, investigating if our findings generalize to other tasks, adjacent to query simulation, such as (next) utterance prediction (e.g., Sim4IA 2025 Task~B), and how these or alternative measures could be employed to assess similarity across entire interaction sessions is another interesting direction for future work. Finally, the taxonomy could be extended with additional facets and measures, such as query performance prediction measures that propose an alternative way to predict query performance without requiring relevance judgments.

%% file: ecir2026-querysim-appendix.tex
\section*{Appendix}
\label{sec:exp:library}

\subsection*{Software Library}
We implement a Python library to compute the measures listed in the taxonomy and support one-to-one and one-to-many comparisons.\footnote{\url{https://github.com/irgroup/query_sim_validation}} It provides a unified framework for validating query simulation approaches, and was used to compute the comparisons in Section~\ref{sec:results}. The library includes implementation the following measures:

\begin{itemize}
    \item Basic query statistics like length, number of terms, or named entities 
    \item Flesch-Kincaid grade scores
    \item Type-Token Ratio (TTR)
    \item Jaccard similarity
    \item Cosine similarity for different embedding models
    \item BERT score
    \item WordNet-based similarity
    \item Various retrieval performance metrics 
    \item SERP overlap based on Jaccard index and RBO
\end{itemize}

The library compares both original and simulated sessions, encoded in JSON. We introduce a session data model that includes a session ID, an ID, interactions, and, optionally, a rank for a simulated session. Note that original and simulated sessions must contain matching session IDs for comparison. An interaction comprises a query, a search engine result page, and, if available, clicked document IDs.

We provide a script to compute some of the measures implemented as an example of how to operate the library. Upon execution, the results are provided in JSONL format, where each line corresponds to one simulator and contains, for each measure, a list of all calculated values in ranking order. These results can then be further processed for descriptive statistics or statistical testing.

%% file: ecir2026-querysim.bib
@inproceedings{Alaofi:2023:SIGIR,
	title        = {Can Generative {LLMs} Create Query Variants for Test Collections? {A}n Exploratory Study},
	author       = {Alaofi, Marwah and Gallagher, Luke and Sanderson, Mark and Scholer, Falk and Thomas, Paul},
	year         = {2023},
	booktitle    = {Proceedings of the 46th International ACM SIGIR Conference on Research and Development in Information Retrieval},
	series       = {SIGIR '23},
	pages        = {1869--1873}
}

@inproceedings{Alaofi:2025:ICTIR,
	title        = {Demographically-Inspired Query Variants Using an {LLM}},
	author       = {Alaofi, Marwah and Ferro, Nicola and Thomas, Paul and Scholer, Falk and Sanderson, Mark},
	year         = {2025},
	booktitle    = {Proceedings of the 2025 International ACM SIGIR Conference on Innovative Concepts and Theories in Information Retrieval (ICTIR)},
	series       = {ICTIR '25},
	pages        = {390--400}
}

@inproceedings{Allan:2017:TREC,
	title        = {{TREC} 2017 Common Core Track Overview},
	author       = {Allan, James and Harman, Donna and Kanoulas, Evangelos and Li, Dan and Van Gysel, Christophe and Voorhees, Ellen M},
	year         = {2017},
	booktitle    = {TREC},
	series       = {TREC '17}
}

@inproceedings{Azzopardi:2007:SIGIR,
	title        = {Building simulated queries for known-item topics: an analysis using six european languages},
	author       = {Azzopardi, Leif and de Rijke, Maarten and Balog, Krisztian},
	year         = {2007},
	booktitle    = {Proceedings of the 30th Annual International ACM SIGIR Conference on Research and Development in Information Retrieval},
	series       = {SIGIR '07},
	pages        = {455--462}
}

@inproceedings{Bailey:2016:SIGIR,
	title        = {{UQV100}: A Test Collection with Query Variability},
	author       = {Bailey, Peter and Moffat, Alistair and Scholer, Falk and Thomas, Paul},
	year         = {2016},
	booktitle    = {Proceedings of the 39th International ACM SIGIR Conference on Research and Development in Information Retrieval},
	series       = {SIGIR '16},
	pages        = {725--728}
}

@article{Bajaj:2018:arXiv,
	title        = {{MS MARCO}: A Human Generated MAchine Reading COmprehension Dataset},
	author       = {Payal Bajaj and Daniel Campos and Nick Craswell and Li Deng and Jianfeng Gao and Xiaodong Liu and Rangan Majumder and Andrew McNamara and Bhaskar Mitra and Tri Nguyen and Mir Rosenberg and Xia Song and Alina Stoica and Saurabh Tiwary and Tong Wang},
	year         = {2018},
	journal      = {cs.CL/1611.09268}
}

@article{Balog:2024:FnTIR,
	title        = {User Simulation for Evaluating Information Access Systems},
	author       = {Krisztian Balog and ChengXiang Zhai},
	year         = {2024},
	journal      = {Foundations and Trends in Information Retrieval},
	volume       = {18},
	number       = {1-2},
	pages        = {1--261},
	issn         = {1554-0669}
}

@inproceedings{Ben:2005:ECIR,
	title        = {Term Frequency Normalisation Tuning for BM25 and DFR Models},
	author       = {He, Ben and Ounis, Iadh},
	year         = {2005},
	booktitle    = {Advances in Information Retrieval},
	series       = {ECIR '05},
	pages        = {200--214}
}

@inproceedings{Berendsen:2012:CLEF,
	title        = {Generating Pseudo Test Collections for Learning to Rank Scientific Articles},
	author       = {Berendsen, Richard and Tsagkias, Manos and de Rijke, Maarten and Meij, Edgar},
	year         = {2012},
	booktitle    = {Information Access Evaluation. Multilinguality, Multimodality, and Visual Analytics},
	series       = {CLEF '12},
	pages        = {42--53}
}

@inproceedings{Breuer:2022:ECIR,
	title        = {Validating Simulations of User Query Variants},
	author       = {Breuer, Timo and Fuhr, Norbert and Schaer, Philipp},
	year         = {2022},
	booktitle    = {Advances in Information Retrieval},
	series       = {ECIR '22},
	pages        = {80--94}
}

@article{Breuer:2025:SIGIRForum,
	title        = {Report on the 1st Workshop on Simulations for Information Access ({Sim4IA} 2024) at {SIGIR} 2024},
	author       = {Breuer, Timo and Kreutz, Christin Katharina and Fuhr, Norbert and Balog, Krisztian and Schaer, Philipp and Bernard, Nolwenn and Frommholz, Ingo and Gohsen, Marcel and Ji, Kaixin and Jones, Gareth J. F. and Keller, J\"{u}ri and Liu, Jiqun and Mladenov, Martin and Pasi, Gabriella and Trippas, Johanne and Wang, Xi and Zerhoudi, Saber and Zhai, ChengXiang},
	year         = {2025},
	journal      = {SIGIR Forum},
	volume       = {58},
	number       = {2},
	pages        = {1--14}
}

@inproceedings{Cai:2009:IEEE,
	title        = {Ontology Driven Semantic Search over Structure P2P Network},
	author       = {Cai, Jinqing and Shao, Xiuli and Ma, Wenhui},
	year         = {2009},
	booktitle    = {2009 Ninth International Conference on Hybrid Intelligent Systems},
	pages        = {29--34},
	series = {HIS '09}
}

@inproceedings{Carterette:2015:ICTIR,
	title        = {Dynamic Test Collections for Retrieval Evaluation},
	author       = {Carterette, Ben and Bah, Ashraf and Zengin, Mustafa},
	year         = {2015},
	booktitle    = {Proceedings of the 2015 International Conference on The Theory of Information Retrieval},
	series       = {ICTIR '15},
	pages        = {91--100}
}

@book{Chuklin:2015:book,
	title        = {Click Models for Web Search},
	author       = {Chuklin, Aleksandr and Markov, Ilya  and de Rijke, Maarten},
	year         = {2015},
	publisher    = {Springer Cham},
	series       = {Synthesis Lectures on Information Concepts, Retrieval, and Services}
}

@inproceedings{Elsweiler:2011:SIGIR,
	title        = {Seeding simulated queries with user-study data for personal search evaluation},
	author       = {Elsweiler, David and Losada, David E. and Toucedo, Jos\'{e} C. and Fernandez, Ronald T.},
	year         = {2011},
	booktitle    = {Proceedings of the 34th International ACM SIGIR Conference on Research and Development in Information Retrieval},
	series       = {SIGIR '11},
	pages        = {25--34},
}

@inproceedings{Elsweiler:2025:SIGIR,
	title        = {Query Smarter, Trust Better? Exploring Search Behaviours for Verifying News Accuracy},
	author       = {Elsweiler, David and Ateia, Samy and Bink, Markus and Donabauer, Gregor and Fern\'{a}ndez Pichel, Marcos and Frummet, Alexander and Kruschwitz, Udo and Losada, David E. and Ludwig, Bernd and Meyer, Selina and Pascual Presa, Noel},
	year         = {2025},
	booktitle    = {Proceedings of the 48th International ACM SIGIR Conference on Research and Development in Information Retrieval},
	series       = {SIGIR '25},
	pages        = {515--526}
}

@inproceedings{Engelmann:2024:ECIR,
	title        = {Context-Driven Interactive Query Simulations Based on Generative Large Language Models},
	author       = {Engelmann, Bj{\"o}rn and Breuer, Timo and Friese, Jana Isabelle and Schaer, Philipp and Fuhr, Norbert},
	year         = {2024},
	booktitle    = {Advances in Information Retrieval},
	series       = {ECIR '24},
	pages        = {173--188}
}

@inproceedings{Erbacher:2022:SIGIR,
	title        = {Interactive Query Clarification and Refinement via User Simulation},
	author       = {Erbacher, Pierre and Denoyer, Ludovic and Soulier, Laure},
	year         = {2022},
	booktitle    = {Proceedings of the 45th International ACM SIGIR Conference on Research and Development in Information Retrieval},
	series       = {SIGIR '22},
	pages        = {2420--2425}
}

@inproceedings{Gunther:2021:Sim4IR,
	title        = {Assessing query suggestions for search session simulation},
	author       = {G{\"u}nther, Sebastian and Hagen, Matthias},
	year         = {2021},
	booktitle    = {Causality in Search and Recommendation (CSR) and Simulation of Information Retrieval Evaluation (Sim4IR) workshops at SIGIR 2021},
	series       = {CSR-Sim4IR '21}
}

@article{Gusenbauer:2020:ResSynMeth,
	title        = {Which Academic Search Systems are Suitable for Systematic Reviews or Meta-analyses? Evaluating Retrieval Qualities of Google Scholar, PubMed, and 26 Other Resources},
	author       = {Gusenbauer, Michael and Haddaway, Neal R.},
	year         = {2020},
	journal      = {Research Synthesis Methods},
	volume       = {11},
	number       = {2},
	pages        = {181--217}
}

@inproceedings{He:2025:SIGIR,
	title        = {{Tip of the Tongue} Query Elicitation for Simulated Evaluation},
	author       = {He, Yifan and Kim, To Eun and Diaz, Fernando and Arguello, Jaime and Mitra, Bhaskar},
	year         = {2025},
	booktitle    = {Proceedings of the 48th International ACM SIGIR Conference on Research and Development in Information Retrieval},
	series       = {SIGIR '25},
	pages        = {3398--3407}
}

@inproceedings{Huurnink:2010:CLEF,
	title        = {Validating Query Simulators: An Experiment Using Commercial Searches and Purchases},
	author       = {Huurnink, Bouke and Hofmann, Katja and de Rijke, Maarten and Bron, Marc},
	year         = {2010},
	booktitle    = {Multilingual and Multimodal Information Access Evaluation},
	series       = {CLEF '10},
	pages        = {40--51}
}

@inproceedings{Kruff:2026:ECIR,
	title        = {Sim4IA-Bench: A User Simulation Benchmark Suite for Next Query and Utterance Prediction},
	author       = {Kruff, Andreas Konstantin and Kreutz, Christin Katharina and Breuer, Timo and Schaer, Philipp and Balog, Krisztian},
	year         = {2026},
	booktitle       = {Advances in Information Retrieval},
    series = {ECIR '26}
}

@inproceedings{Labhishetty:2021:SIGIR,
	title        = {An Exploration of Tester-based Evaluation of User Simulators for Comparing Interactive Retrieval Systems.},
	author       = {Labhishetty, Sahiti and Zhai, Chengxiang},
	year         = {2021},
	booktitle    = {Proceedings of the 44th International ACM SIGIR Conference on Research and Development in Information Retrieval},
	series       = {SIGIR '21},
	pages        = {1598--1602},
}

@inproceedings{Labhishetty:2022:ECIR,
	title        = {{RATE}: A Reliability-Aware Tester-Based Evaluation Framework of User Simulators},
	author       = {Labhishetty, Sahiti and Zhai, ChengXiang},
	year         = {2022},
	booktitle    = {Advances in Information Retrieval},
	series       = {ECIR '22},
	pages        = {336--350}
}

@inproceedings{Labhishetty:2022:ICTIR,
	title        = {{PRE}: A Precision-Recall-Effort Optimization Framework for Query Simulation},
	author       = {Labhishetty, Sahiti and Zhai, ChengXiang},
	year         = {2022},
	booktitle    = {Proceedings of the 2022 ACM SIGIR International Conference on Theory of Information Retrieval},
	series       = {ICTIR '22},
	pages        = {51--60}
}

@inproceedings{Morrison:2011:IEEE,
	title        = {Query log simulation for long-term learning in image retrieval},
	author       = {Morrison, Donn and Marchand-Maillet, Stéphane and Bruno, {\'E}ric},
	year         = {2011},
	booktitle    = {2011 9th International Workshop on Content-Based Multimedia Indexing},
	pages        = {55--60},
    series = {CBMI '11}
}

@inproceedings{Rahmani:2025:CIKM,
	title        = {Towards Understanding Bias in Synthetic Data for Evaluation},
	author       = {Rahmani, Hossein A. and Ramineni, Varsha and Yilmaz, Emine and Craswell, Nick and Mitra, Bhaskar},
	year         = {2025},
	booktitle    = {Proceedings of the 34th ACM International Conference on Information and Knowledge Management},
	series       = {CIKM '25},
	pages        = {5166--5170}
}

@inproceedings{Schaer:2025:SIGIR,
	title        = {Second {SIGIR} Workshop on Simulations for Information Access ({Sim4IA} 2025)},
	author       = {Schaer, Philipp and Kreutz, Christin Katharina and Balog, Krisztian and Breuer, Timo and Kruff, Andreas Konstantin},
	year         = {2025},
	booktitle    = {Proceedings of the 48th International ACM SIGIR Conference on Research and Development in Information Retrieval},
	series       = {SIGIR '25},
	pages        = {4172--4175}
}

@inproceedings{Sinha:2024:ECIR,
	title        = {Exploring the Nexus Between Retrievability and Query Generation Strategies},
	author       = {Sinha, Aman and Mall, Priyanshu Raj and Roy, Dwaipayan},
	year         = {2024},
	booktitle    = {Advances in Information Retrieval},
	series       = {ECIR '24},
	pages        = {177--193}
}

@inproceedings{Traub:2016:JCDL,
	title        = {Querylog-based Assessment of Retrievability Bias in a Large Newspaper Corpus},
	author       = {Traub, Myriam C. and Samar, Thaer and van Ossenbruggen, Jacco and He, Jiyin and de Vries, Arjen and Hardman, Lynda},
	year         = {2016},
	booktitle    = {Proceedings of the 16th ACM/IEEE-CS on Joint Conference on Digital Libraries},
	series       = {JCDL '16},
	pages        = {7--16}
}

@book{Zeigler:2019:book,
	title        = {Theory of Modeling and Simulation: Discrete Event \& Iterative System Computational Foundations, Third Edition},
	author       = {Zeigler, Bernard P. and Muzy, Alexandre and Kofman, Ernesto},
	year         = {2019},
	publisher    = {Elsevier}
}

@inproceedings{Zendel:2025:CIKM,
	title        = {A Comparative Analysis of Linguistic and Retrieval Diversity in LLM-Generated Search Queries},
	author       = {Zendel, Oleg and Al Lawati, Sara Fahad Dawood and Rashidi, Lida and Scholer, Falk and Sanderson, Mark},
	year         = {2025},
	booktitle    = {Proceedings of the 34th ACM International Conference on Information and Knowledge Management},
	series       = {CIKM '25},
	pages        = {4014--4023}
}

@inproceedings{Zerhoudi:2022:TPDL,
	title        = {Simulating User Querying Behavior Using Embedding Space Alignment},
	author       = {Zerhoudi, Saber and Granitzer, Michael},
	year         = {2022},
	booktitle    = {Linking Theory and Practice of Digital Libraries},
	series       = {TPDL '22},
	pages        = {386--394}
}

@inproceedings{Zhang:2025:SIGIR,
	title        = {Exploring Human-Like Thinking in Search Simulations with Large Language Models},
	author       = {Zhang, Erhan and Wang, Xingzhu and Gong, Peiyuan and Yang, Zixuan and Mao, Jiaxin},
	year         = {2025},
	booktitle    = {Proceedings of the 48th International ACM SIGIR Conference on Research and Development in Information Retrieval},
	series       = {SIGIR '25},
	pages        = {2669--2673}
}
